\documentclass[prb,12pt,notitlepage,tightenlines,superscriptaddress]{revtex4}
\pdfoutput=1

\usepackage{amsmath}
\usepackage{amsfonts}
\usepackage{bm}
\usepackage{euscript}
\usepackage{graphicx}
\usepackage{wasysym}
\usepackage{color}
\usepackage{subfigure}
\usepackage{wrapfig}

\graphicspath{{figs/}}

\newcommand{\mathnotation}[2]{\newcommand{#1}{\ensuremath{#2}}}

\mathnotation{\pd}{\partial}           % Partial derivative
\mathnotation{\ee}{{\mathrm e}}        % e
\mathnotation{\imi}{\mathrm{i}}        % i
\mathnotation{\ldef}{\mathrel{\raisebox{.069ex}{:}\!\!=}}% Left define
\mathnotation{\rdef}{\mathrel{=\!\!\raisebox{.069ex}{:}}}% Right define
\mathnotation{\surf}{S}
\mathnotation{\MCG}{\textrm{MCG}}
\mathnotation{\trace}{\mathrm{Tr}}
\mathnotation{\Ind}{\mathrm{Ind}}
\mathnotation{\Fix}{\mathrm{Fix}}
\mathnotation{\fp}{p}
\mathnotation{\pr}{N}

\newcommand{\madisonaddress}{Department of Mathematics, University of
  Wisconsin -- Madison, USA}

\begin{document}

\title{The cat's cradle, stirring, and topological complexity}

\author{Jean-Luc Thiffeault}
\email{jeanluc@math.wisc.edu}
\affiliation{\madisonaddress}
\author{Erwan Lanneau}
\email{erwan.lanneau@cpt.univ-mrs.fr}
\address{
Centre de Physique Th\'eorique,
Universit\'e du Sud Toulon-Var and
F\'ed\'eration de Recherches des Unit\'es de
Math\'ematiques de Marseille,
Luminy, France}
\author{Sarah Matz}
\email{matz@math.wisc.edu}
\affiliation{\madisonaddress}

\begin{abstract}
  There are several physical situations in which the `tangling' of a
  loop is relevant: the game of cat's cradle is a simple example, but
  a more important application involves the stirring of a fluid by
  rods.  Here we discuss how elementary topology constrains the types
  of mappings that can occur on a surface, for example when the
  surface is the domain of a two-dimensional fluid.
\end{abstract}

\maketitle

\noindent
\textit{This is an article for Dynamical Systems Magazine
  (http://www.dynamicalsystems.org/)}

\bigskip

%\section{Introduction}
%\label{sec:intro}

\begin{wrapfigure}{R}{.52\textwidth}
  \centerline{\includegraphics[width=.5\textwidth]{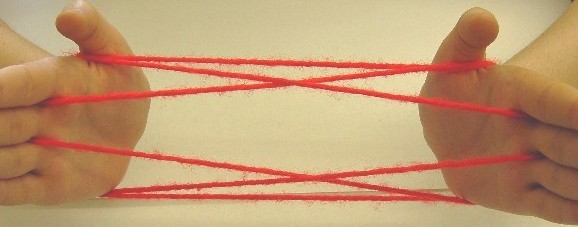}}
  \caption{The \emph{cat's cradle}.}
  \label{fig:cat}
\end{wrapfigure}

In the children's game of \emph{cat's cradle}, a loop of string is
first anchored around both hands.  The fingers of each hand are
then inserted into the opposite hand's loop and pulled apart,
resulting in the situation shown in Figure~\ref{fig:cat}.  The motion
of the fingers allows us to predict in advance just how long a piece
of string we will need: the more intricate the motion, the more string
is required.
In fact the length of string is not a bad first definition of `topological
complexity,' and we will see how this simple concept can be applied to
the study of dynamical systems, particularly in two dimensions.  The
game of cat's cradle also suggests a connection between topology and
combinatorics: after all, the number of distinct cat's cradle shapes
that can be made is essentially a difficult combinatorial problem.
Throughout this short note, we will explore the interplay between
topology and combinatorics, and see what it reveals about dynamical
systems on surfaces.

\begin{wrapfigure}[16]{L}{.4\textwidth}
  \centerline{\includegraphics[width=.38\textwidth]{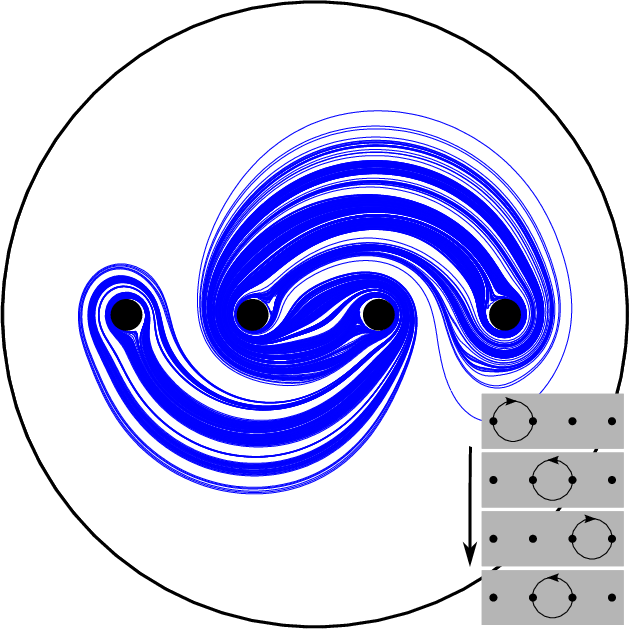}}
  \caption{A stirring device with four rods.  The inset shows the
    sequence of rod motions, and the blue line is a blob of dye that
    was filamented by the rods. (From
    Ref.~\onlinecite{Thiffeault2008b}.)}
  \label{fig:s1s-2s3s-2}
\end{wrapfigure}

On the practical side of things, this sort of approach to topological
complexity is useful when studying stirring
devices.~\cite{Boyland2000,Thiffeault2006,Thiffeault2008b}
Figure~\ref{fig:s1s-2s3s-2} shows a top-down view of such a device:
the black disks are stirring rods immersed in a two-dimensional white
fluid, assumed very viscous.  The rods are moved following the
prescription shown in the inset.  The result is that an initial loop
gets wrapped around the rods in a topologically complex manner, and
is stretched exponentially (blue line).  This is very good for the ultimate goal,
which is to enhance the chemical reaction rate or blending of
solutes: the blue line represents the contact area between the two
reagents.  The reaction rate is limited by the growth of this contact
area, so having it grow exponentially is desirable.

\section*{Two handles are better than one}

\begin{wrapfigure}{R}{.4\textwidth}
  \centerline{\includegraphics[width=.38\textwidth]{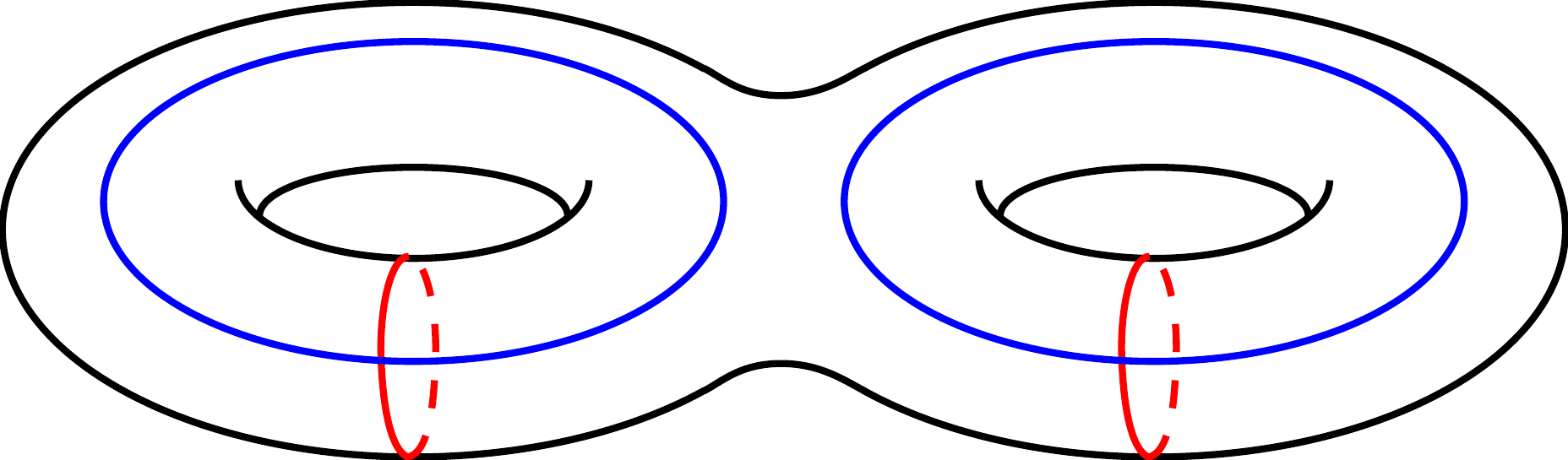}}
  \caption{The double-torus, $\surf_2$.}
  \label{fig:2torus}
\end{wrapfigure}

Figure~\ref{fig:2torus} shows the double-torus~$\surf_2$, beloved of
low-dimensional topologists.  The subscript 2 on $\surf_2$ stands for
the number of handles, or \emph{genus}.  Though it isn't
encountered very much in the real world, it is a surface that's
complex enough to be interesting, unlike the humble ordinary torus
$\surf_1$ which is a bit too trivial.  We are interested in
transformations of a surface $\surf$, i.e., continuous mappings
of~$\surf$ to itself.  Such a mapping, if also invertible, is called a
homeomorphism of~$\surf$.  (We assume that the surface is orientable,
and that the mapping preserves this orientation.)  Now in topology one
only cares about gross features, so we assume that two mappings are
equivalent if they can be continuously deformed into each other: they
are then called \emph{isotopic}.  An equivalence class of mappings
under isotopy is called an \emph{isotopy class}.  The set of all such
classes forms the \emph{mapping class group} $\MCG(\surf)$ of~$\surf$,
with the group operation obtained from composition of mappings.

\begin{figure}
  \centerline{\includegraphics[width=.43\textwidth]{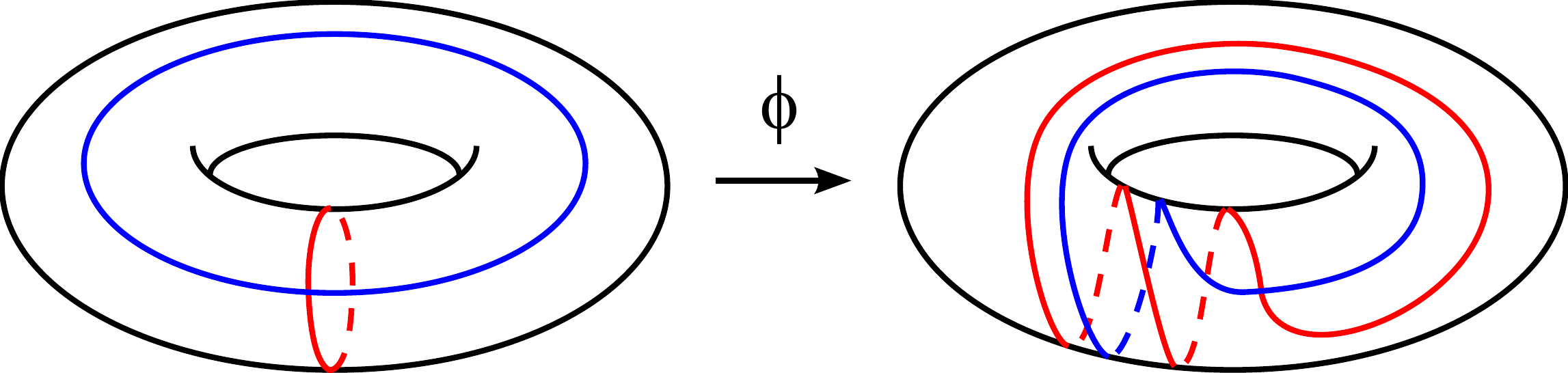}}
  \caption{The action on loops of a mapping of the torus (their
    orientation is not shown).}
  \label{fig:torus}
\end{figure}

The mapping class group of the torus $\surf_1$ is particularly simple:
after we allow for isotopy, all that matters is how many times a
mapping $\phi:\surf_1\rightarrow\surf_1$ has caused us to wind around
each of the two periodic directions of the torus.  This winding is
best illustrated by looking at the action of a mapping on two
nontrivial disjoint closed curves drawn on the surface, as in
Figure~\ref{fig:torus}.  The red loop around the thin direction of the
torus is labeled $\begin{pmatrix}1\\0\end{pmatrix}$, the blue loop
$\begin{pmatrix}0\\1\end{pmatrix}$.  The mapping in
Figure~\ref{fig:torus} can be written as
$\begin{pmatrix}2&1\\1&1\end{pmatrix}$: its linear action transforms
the red loop to $\begin{pmatrix}2\\1\end{pmatrix}$ and the blue loop
to $\begin{pmatrix}1\\1\end{pmatrix}$, which indicates the number of
twists of each image loop around each periodic direction of the torus.
In fact, in general we have
$\MCG(\surf_1)\simeq\mathrm{PSL}_2(\mathbb{Z})$ (the ``projective''
version of the group of invertible two-by-two matrices with
determinant 1), since we can wind in positive as well as negative
directions, and the action of~$\phi$ on these loops is commutative in
the case of the torus.  Note that since the loops initially only
intersect once, then their images only intersects once as well.

According to the celebrated Thurston--Nielsen (TN) classification
theorem,~\cite{Thurston1988} there are only three types of isotopy
classes: finite-order, reducible, and pseudo-Anosov.  A mapping
$\phi:\surf\rightarrow\surf$ is finite-order if~$\phi^k$ is isotopic
to the identity map, for some integer~$k>0$.  The mapping $\phi$ is
reducible if it leaves a set of disjoint closed curves invariant: the
surface is carved up such that the pieces map to each other.
Finite-order mappings are not interesting to us because they are too
simple; reducible ones are not interesting because one can then
reapply the TN theorem to the individual pieces (with boundaries)
until the mapping is no longer reducible.

% Can't get this to wrap properly, so use figure rather than wrapfigure.
%
%\begin{wrapfigure}{L}{.45\textwidth}

It is the third type --- pseudo-Anosov --- that is most dynamically
interesting.  For such a mapping, in addition to leaving no nontrivial
closed curve ``untouched'' (otherwise it would be reducible), the
action on loops causes their length to grow
exponentially.~\cite{endnote1} The asymptotic factor by which the
winding numbers of loops is multiplied (under repeated iteration) is
called the \emph{dilatation}, $\lambda>1$.  Its logarithm is the
\emph{topological entropy}.  The topological entropy is closely
related to the growth rate of loops, as discussed for the cat's cradle
and stirring device examples in the introduction.  The positive
entropy of a pseudo-Anosov mapping tells us that loops will grow
exponentially under repeated iteration, with growth
rate~$\log\lambda$.

For the torus~$\surf_1$, the TN classification is achieved merely by
looking at the trace of an element of
$\MCG(\surf_1)\simeq\mathrm{PSL}_2(\mathbb{Z})$.  If the absolute
value of the trace is greater than 2, then the mapping is
pseudo-Anosov;~\cite{endnote2} If less than 2, finite-order; If equal
to 2, reducible.~\cite{endnote3} One can check this easily by using
the Cayley--Hamilton theorem for two-by-two matrices of unit
determinant.  The mapping of the torus in Figure~\ref{fig:torus}, for
instance, is pseudo-Anosov with~$\lambda\simeq 2.618$.

\section*{Of polynomials and indices}

Now let us return to the double-torus~$\surf_2$.  The mapping class
group in this case is more complicated than for the torus, but we can
get a handle on part of it by considering $\MCG^+(\surf_2)$ --- the
subset of the mapping class group that also leaves invariant a
vector field (as opposed to a line field in the general case).  In
that case the mapping class group can be deduced from the action on
the four loops drawn in Figure~\ref{fig:2torus}, so that
$\MCG^+(\surf_2) \subset \mathrm{SL}_4(\mathbb{Z})$.~\cite{endnote4}
But this time we do not have equality: there are many matrices that do
not correspond to an element of the mapping class group.  But which
ones?

The matrices we want are in the \emph{symplectic group}
$\mathrm{Sp}_{4}(\mathbb{Z})$.~\cite{endnote5} One important feature
of these matrices is that their characteristic polynomial
\begin{equation*}
  P(x) = x^4 + c_1\,x^3+c_2\,x^2+c_3\,x + c_4
\end{equation*}
satisfies~$c_3=c_1$, $c_4=1$, so that the polynomial is
\emph{reciprocal} --- the coefficients %are palindromic and so
read the
same from left to right as from right to left.  The largest root (in
magnitude) of~$P(x)$ is the dilatation $\lambda$; for a pseudo-Anosov
mapping it is always real.  We appeal to the \emph{Lefschetz fixed point
  theorem}, which in our case says
\begin{equation*}
  L(\phi) = 2 - \trace(\phi_*) = \sum_{\fp \in \Fix(\phi)}\Ind(\phi,\fp).
\end{equation*}
Here, $L(\phi)$ is an integer called the Lefschetz number of $\phi$;
$\phi_* \in \mathrm{Sp}_4(\mathbb{Z})$ is a matrix representing the
isotopy class of $\phi$ by looking at its action on the loops in
Figure~\ref{fig:2torus}; $\trace(\phi_*)$ is the trace; $\Fix(\phi)$
is the set of isolated fixed points of~$\phi$; and $\Ind(\phi,\fp)$ is
the \emph{topological index} of $\phi$ at the
\begin{wrapfigure}{R}{.52\textwidth}
  \centerline{\includegraphics[width=.5\textwidth]{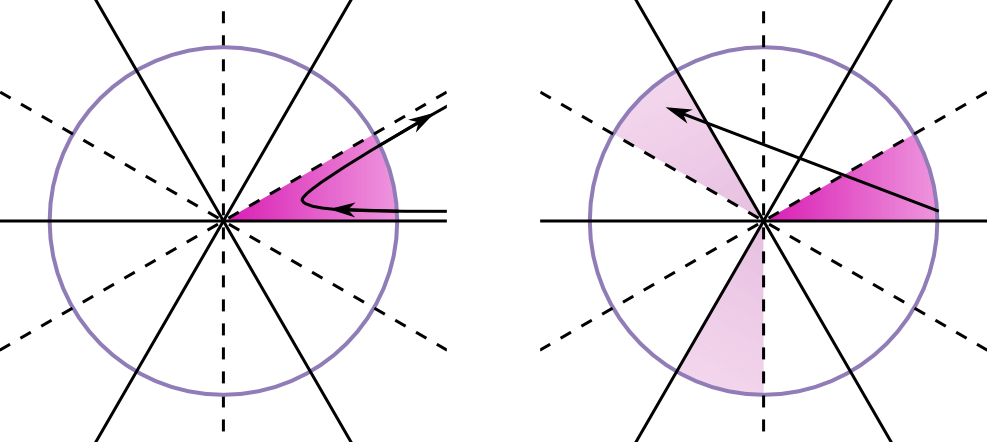}}
  \caption{A singularity with 12 hyperbolic sectors ($\pr=6$).  Each
    sector can map to itself (left, index~$1-6=-5$) or to one
    of two other sectors (right, index~$+1$). The index is defined as
    the number of turns of a vector joining $x$ to $\phi(x)$ as $x$
    travels counterclockwise around a small circle.}
  \label{fig:index_sing6}
\end{wrapfigure}
isolated fixed point $\fp$.  The topological index is given by the signed
number of turns of a vector joining a point~$x$ and its
image~$\phi(x)$ as $x$ travels counterclockwise around a small circle
enclosing the singularity (see Figure~\ref{fig:index_sing6}).  The index
can only assume two simple values: it is~$1-\pr$ if~$\phi$ fixes
the $2\pr$ `hyperbolic sectors' around a fixed point, and it
is~$+1$ (independent of~$\pr$) if~$\phi$ permutes the
sectors.~\cite{endnote6} This is due to the severe restrictions
imposed by continuity and
orientation-preservation.~\cite{LanneauThiffeault}

Now comes the first crucial ingredient: the Cayley--Hamilton theorem
and Newton's formulas~\cite{Uspensky} tell us how to find
$\trace(\phi_*^k)$ recursively by looking at the polynomial~$P(x)$:
\begin{equation*}
  \trace(\phi_*^k) = -\sum_{m=1}^{k-1} c_m \trace(\phi_*^{k-m}) - kc_k,
\end{equation*}
where~$c_k=0$ for~$k>n$.  Hence, we can calculate the Lefschetz
numbers of all powers of~$\phi$ directly from the polynomial.  For
example, for the polynomial~\cite{ChoHam2008}
\begin{equation*}
  P(x) = x^4 - x^3 - x^2 - x + 1
\end{equation*}
with dilatation~$\lambda\simeq1.72208$, we find
\begin{equation*}
%L(\phi^0)=-2,\ \
L(\phi)=1,\ \ L(\phi^2)=-1,\ \ L(\phi^3)=-5,\ \ L(\phi^4)=-5,\ \
L(\phi^5)=-14,\ \ L(\phi^6)=-25\dots.
\end{equation*}
%{1, -1, -5, -5, -14, -25, -41, -77, -131, -226, -395, -677, -1169, -2017, -3470, -5981, -10301, -17737, -30551, -52610}

The second crucial ingredient is that the index $\Ind(\phi,\fp)$ at a
fixed point $\fp$ is heavily constrained.  Around a fixed point, the
vector field that remains invariant under~$\phi$ can have a
singularity, near which the mapping's action can be divided
into~$2\pr$ hyperbolic sectors as in Figure~\ref{fig:index_sing6},
with~$\pr$ even.  Assuming the polynomial~$P(x)$ has a positive dominant
root, then as mentioned above there are two cases:
$\Ind(\phi,\fp)=1-\pr<0$ if $\phi$ fixes the sectors at~$\fp$ (left in
Figure~\ref{fig:index_sing6}); $\Ind(\phi,\fp)=1$ if it permutes the
sectors (right in Figure~\ref{fig:index_sing6}). For technical
reasons,~\cite{LanneauThiffeault} a permutation involves only $\pr/2$
of the sectors.

Let's see what we can glean from this for the case of the polynomial
above: on a surface of genus two, the Euler--Poincar\'e formula
dictates
\begin{equation*}
  \sum_{\fp \in \Fix(\phi)} (2-\pr_\fp) = 2\chi_2 = -4
\end{equation*}
where~$\chi_2=-2$ is the Euler characteristic of the
surface~$\surf_2$, and~$2\pr_\fp$ is the number of hyperbolic sectors
at a fixed point~$\fp$.

Note first that if~$\pr = 2$ the fixed point contributes nothing to
the left side (from now on we drop the~$\fp$ subscript
from~$\pr_\fp$), so according to the formula we can have any number of
fixed points of this type. These are called \emph{regular fixed
  points}, and can only have a index of~$-1$ ($\pr/2 = 1$, so the only
permutation of sectors is the identity). Then, ignoring regular fixed
points, the only possibilities for singularities are either one
with~$\pr=6$, or two with~$\pr=4$ (recall~$\pr$ must be even).

First consider the case of two singularities
with~$\pr=4$. Since~$L(\phi)=1$, at least one of the singularities
must be fixed with its sectors permuted, since this is the only way to
get a positive index. When $\pr=4$ the permutation involves only two
sectors, so $\phi^2$ will fix the sectors and the singularity will
have an index of $1-\pr = -3$.  However, the other singularity can
contribute at most~$+1$ to $L(\phi^2)$, so this is incompatible with
the Lefschetz number $L(\phi^2)=-1$. We conclude that this polynomial
cannot correspond to a pseudo-Anosov mapping involving two singularities
with~$\pr=4$.

Now consider the second case: one singularity with~$\pr=6$.
Since~$L(\phi)=1$, the sectors must be permuted to get a positive
index.  When~$\pr=6$ the permutation involves~$\pr/2 = 3$ sectors
(Figure~\ref{fig:index_sing6}).  Therefore,~$\phi^2$ also permutes the
sectors (so that the singularity contributes~$+1$ to the Lefschetz
number) and $\phi^3$ fixes the sectors (so the singularity
contributes~$1-\pr = -5$).  Then, in order to match the Lefschetz
number~$L(\phi^2) = -1$ there must be two regular fixed points, which
together make a regular period-2 orbit.  Regular fixed points must
appear at a multiple of their period (that is, a period-$k$ orbit
becomes~$k$ fixed points of $\phi^k, \phi^{2k}$, etc.).  Hence, we
conclude that a period-2 regular orbit exists, so that $L(\phi^2) =
1-2\cdot(-1) = -1$, as required.

Next, for the third iterate~$\phi^3$, the Lefschetz number
is~$L(\phi^3)=-5$: this is fine since the 3 sectors of our~$\pr=6$
singularity are now unpermuted, and the index is thus~$1-\pr=1-6=-5$.
This means that there are no regular orbits of period 3.  For the
fourth iterate, we have~$L(\phi^4)=-5$ again, but this time our main
singularity has its sectors permuted, so its index is~$+1$.  Our
earlier period-2 orbits appear at this iterate and contribute~$-2$,
and that leaves a deficit of~$-4$, which is accounted for by invoking
a single period-4 orbit.

Let's look at one final iterate: for the fifth one~$L(\phi^5)=-14$, so
something big is happening.  Our main singularity with~$\pr=6$ still
has its sectors permuted, so its contribution is~$+1$.  No other
periodic orbits that we've encountered before can contribute, since 5
is a prime iterate.  Hence, the deficit of~$-14-(+1)=-15$ must be
entirely made up by \emph{three} period-5 orbits.

So far the tally is (i) a single fixed point with~$\pr=6$; (ii) a
regular ($\pr=2$) period-2 orbit; (iii) a regular period-4 orbit; and (iv)
three regular period-5 orbits.  Of course, we can keep going, but we
have made our point: the list of Lefschetz numbers allows us to deduce
which periodic orbits must appear at each iterate.  Moreover, it is
clear that the list is not arbitrary: there are many sequences of
Lefschetz numbers that would be inconsistent, in the sense that no
compatible periodic-orbit structure exists.  What is remarkable is that all
this information can be obtained from a simple polynomial.

\section*{The littlest pseudo-Anosov}

The tools discussed so far allow us to answer an interesting question:
for a given surface of genus~$g$, what is the `simplest' pseudo-Anosov
mapping?  By simplest, we mean the one with the least
dilatation~$\lambda$, which we call~$\delta^+_g$ (the `+' superscript
corresponds to~$\MCG^+$).  This is a question that has occupied some
low-dimensional topologists for 20 years, ever since Penner first
posed it.  Originally it wasn't clear how this `minimizer' should
behave with the genus, but better and better examples were
constructed, and currently the best general bound stands
at~$\delta^+_g \lesssim (2+\sqrt{3})^{1/g}$, implying that the
minimizer converges to unity for large genus.~\cite{Hironaka2006} In
fact, this bound is sharp for~$g=1$~\cite{Series1985} and
$2$,~\cite{Zhirov1995,Song2002,Ham2007,ChoHam2008,LanneauThiffeault}
but explicit pseudo-Anosov mappings have been found with dilatation
below the bound for~$3\le g\le
5$.~\cite{LanneauThiffeault,Leininger2004} In genus 5, the
pseudo-Anosov mapping associated with~$\delta^+_5$ has the mysterious
Lehmer's number~$\simeq1.17628$ as a
dilatation.~\cite{Leininger2004,Hironaka_WhatIs}  Moreover, the minimum
dilatations for genus 2 to 5 are all \emph{Salem numbers}: these are
roots of polynomials with only one eigenvalue outside the unit circle
in the complex plane.~\cite{Boyd1980}  Nobody knows what
happens in general for higher genus.

In Ref.~\onlinecite{LanneauThiffeault} we used a very direct method to
prove, amongst others things, the minimality
of the Salem numbers~$\delta^+_3\simeq1.40127$,~$\delta^+_4\simeq1.28064$,
and~$\delta^+_5\simeq1.17628$.  Here, we illustrate the method for genus 3:
Start with the polynomial associated with the candidate pseudo-Anosov
mapping, whose dilatation is believed to give~$\delta^+_3$, in this
case
\begin{equation*}
  P(x) = x^6 - x^4 - x^3 - x^2 + 1
\end{equation*}
with dilatation~$\simeq1.40127$.  Now how many degree 6, reciprocal,
integer-coefficient polynomials have a largest real root that is less
than this dilatation?  A simple computer search tells us that the list
is very small: there is only one, which can be factored as $(x^3 -
x - 1)(x^3 + x^2 - 1)$, with Lefschetz numbers 
% {3, -1, -3, 3, -7, -1, -4, -13, -3, -21}
\begin{equation*}
L(\phi)=3,\ \ L(\phi^2)=-1,\ \ L(\phi^3)=-3,\ \ L(\phi^4)=3,\ \
L(\phi^5)=-7,\ \ L(\phi^6)=-1\dots
\end{equation*}
where~$\phi$ is the hypothetical pseudo-Anosov mapping.  Since there
are a finite number of possible singularities (because of the
Euler--Poincar\'e formula with~$\chi_3=-4$), we can check if this
sequence of Lefschetz numbers is possible.  Because~$L(\phi)=3$,
$\phi$ must fix at least three singularities.  For instance, assume
that~$\phi$ fixes one singularity with~$\pr=6$, permuting its sectors
as in Figure~\ref{fig:index_sing6}, and two with~$\pr=4$ with
their~$\pr/2$ sectors permuted.  Then~$\phi^2$ will fix the sectors of
the two singularities with~$\pr=4$, and each singularity will
contribute~$1-4=-3$ to the Lefschetz number, for a total of~$-6$.  But
from above,~$L(\phi^2)=-1$, which is not possible: there is no way to
make the Lefschetz number more positive by adding regular orbits,
since they contribute negatively.  Hence,~$\phi$ cannot fix these
three types of singularities.  There is only one other possible case
--- four singularities with~$\pr=4$ --- which we can rule out as well.
We conclude that the polynomial $(x^3 - x - 1)(x^3 + x^2 - 1)$ is not
associated with any pseudo-Anosov mapping.  Thus, we have the
mimimizer~$\delta^+_3\simeq1.40127$.  In order to truly complete the
proof, one needs to explicitly construct the action of this
pseudo-Anosov mapping on the surface (see
Ref.~\onlinecite{LanneauThiffeault}).~\cite{endnote6}

In the end, is this numerological exercise worth it?  Well, it
certainly simplifies the study of pseudo-Anosov mappings on surfaces.
In fact, even these abstract surfaces of genus~$g$ have practical
significance: pseudo-Anosov mappings of disks, which can be associated
to a stirring device such as that in Figure~\ref{fig:s1s-2s3s-2}, can
be lifted to surfaces of genus~$g$ by a `branched cover.'  For such
applications, the biggest challenge is to develop practical tools
(such as \emph{train tracks}~\cite{PennerHarer}) that can handle
singularities with an \emph{odd} number of sectors, which are
mathematically more challenging.

\subsection*{Acknowledgments}

We thank Phil Boyland for helpful comments.
J-LT was partially supported by the Division of Mathematical Sciences
of the US National Science Foundation, under grant DMS-0806821.

%\bibliographystyle{jlt}
%\bibliography{bib/journals_abbrev,bib/articles,endnotes}

\begin{thebibliography}{21}
\newcommand{\enquote}[1]{`#1'}
\providecommand{\natexlab}[1]{#1}
\providecommand{\url}[1]{\texttt{#1}}
\providecommand{\urlprefix}{URL }
\providecommand{\bibinfo}[2]{#2}
\providecommand{\eprint}[2][]{\url{#2}}

\bibitem[{Thiffeault \emph{et~al.}(2008)Thiffeault, Finn, Gouillart, and
  Hall}]{Thiffeault2008b}
\bibinfo{author}{J.-L. Thiffeault}, \bibinfo{author}{M.~D. Finn},
  \bibinfo{author}{E.~Gouillart}, and \bibinfo{author}{T.~Hall},
  \enquote{\bibinfo{title}{Topology of chaotic mixing patterns},}
  \emph{\bibinfo{journal}{Chaos}} \textbf{\bibinfo{volume}{18}},
  \bibinfo{pages}{033123} (\bibinfo{year}{2008}), \eprint{arXiv:0804.2520}.

\bibitem[{Boyland \emph{et~al.}(2000)Boyland, Aref, and Stremler}]{Boyland2000}
\bibinfo{author}{P.~L. Boyland}, \bibinfo{author}{H.~Aref}, and
  \bibinfo{author}{M.~A. Stremler}, \enquote{\bibinfo{title}{Topological fluid
  mechanics of stirring},} \emph{\bibinfo{journal}{J. Fluid Mech.}}
  \textbf{\bibinfo{volume}{403}}, \bibinfo{pages}{277--304}
  (\bibinfo{year}{2000}).

\bibitem[{Thiffeault and Finn(2006)}]{Thiffeault2006}
\bibinfo{author}{J.-L. Thiffeault} and \bibinfo{author}{M.~D. Finn},
  \enquote{\bibinfo{title}{Topology, braids, and mixing in fluids},}
  \emph{\bibinfo{journal}{Phil. Trans. R. Soc. Lond. A}}
  \textbf{\bibinfo{volume}{364}}, \bibinfo{pages}{3251--3266}
  (\bibinfo{year}{2006}).

\bibitem[{Thurston(1988)}]{Thurston1988}
\bibinfo{author}{W.~P. Thurston}, \enquote{\bibinfo{title}{On the geometry and
  dynamics of diffeomorphisms of surfaces},} \emph{\bibinfo{journal}{Bull. Am.
  Math. Soc.}} \textbf{\bibinfo{volume}{19}}, \bibinfo{pages}{417--431}
  (\bibinfo{year}{1988}).

\bibitem[{end(????{\natexlab{a}})}]{endnote1}
 \bibinfo{note}{For the experts, we are avoiding here the
  rigorous definition of pseudo-Anosov mappings in terms of invariant measured
  foliations, hopefully without too much harm.}

\bibitem[{end(????{\natexlab{b}})}]{endnote2}
 \bibinfo{note}{Actually, for the torus the `pseudo' is
  dropped because of the absence of singularities.}

\bibitem[{end(????{\natexlab{c}})}]{endnote3}
 \bibinfo{note}{The reducible case is rather degenerate
  for the torus, since there is an infinity of reducing curves.}

\bibitem[{end(????{\natexlab{d}})}]{endnote4}
 \bibinfo{note}{Such a collection of loops is a basis of
  the first homology group~$H_1(\surf_2,\mathbb{Z})$.  Note that we
  are leaving out a surjective group homomorphism from~$\MCG(\surf_2)$ to
  $\mathrm{Sp}_4(\mathbb{Z})$.}

\bibitem[{end(????{\natexlab{e}})}]{endnote5}
 \bibinfo{note}{The symplectic group consists of matrices
  that preserve a skew-symmetric quadratic form.}

\bibitem[{end(????{\natexlab{f}})}]{endnote6}
 \bibinfo{note}{We are assuming~$P(x)$ has a positive largest root;
   for negative root a parallel argument can be made.}

\bibitem[{Lanneau and Thiffeault(2009)}]{LanneauThiffeault}
\bibinfo{author}{E.~Lanneau} and \bibinfo{author}{J.-L. Thiffeault},
  \enquote{\bibinfo{title}{On the minimum dilatation of pseudo-{A}nosov
  homeomorphisms on surfaces of small genus},}   (\bibinfo{year}{2009}),
  \eprint{arXiv:0905.1302}.

\bibitem[{Uspensky(1963)}]{Uspensky}
\bibinfo{author}{J.~C. Uspensky}, \emph{\bibinfo{title}{Theory of Equations}}
  (\bibinfo{publisher}{McGraw-Hill}, \bibinfo{address}{New York},
  \bibinfo{year}{1963}).

\bibitem[{Cho and Ham(2008)}]{ChoHam2008}
\bibinfo{author}{J.-H. Cho} and \bibinfo{author}{J.-Y. Ham},
  \enquote{\bibinfo{title}{The minimal dilatation of a genus-two surface},}
  \emph{\bibinfo{journal}{Experiment. Math.}}
  \textbf{\bibinfo{volume}{17}}~(\bibinfo{number}{3}),
  \bibinfo{pages}{257--267} (\bibinfo{year}{2008}).

\bibitem[{Hironaka and Kin(2006)}]{Hironaka2006}
\bibinfo{author}{E.~Hironaka} and \bibinfo{author}{E.~Kin},
  \enquote{\bibinfo{title}{A family of pseudo-{A}nosov braids with small
  dilatation},} \emph{\bibinfo{journal}{Algebraic \& Geometric Topology}}
  \textbf{\bibinfo{volume}{6}}, \bibinfo{pages}{699--738}
  (\bibinfo{year}{2006}), \eprint{arXiv:math/0507012}.

\bibitem[{Series(1985)}]{Series1985}
\bibinfo{author}{C.~Series}, \enquote{\bibinfo{title}{The modular surface and
  continued fractions},} \emph{\bibinfo{journal}{J. London Math. Soc. (2)}}
  \textbf{\bibinfo{volume}{31}}~(\bibinfo{number}{1}), \bibinfo{pages}{69--80}
  (\bibinfo{year}{1985}).

\bibitem[{Song \emph{et~al.}(2002)Song, Ko, and Los}]{Song2002}
\bibinfo{author}{W.~T. Song}, \bibinfo{author}{K.~H. Ko}, and
  \bibinfo{author}{J.~E. Los}, \enquote{\bibinfo{title}{Entropies of braids},}
  \emph{\bibinfo{journal}{J. Knot Th. Ramifications}}
  \textbf{\bibinfo{volume}{11}}~(\bibinfo{number}{4}),
  \bibinfo{pages}{647--666} (\bibinfo{year}{2002}).

\bibitem[{Ham and Song(2007)}]{Ham2007}
\bibinfo{author}{J.-Y. Ham} and \bibinfo{author}{W.~T. Song},
  \enquote{\bibinfo{title}{The minimum dilatation of pseudo-{A}nosov
  5-braids},} \emph{\bibinfo{journal}{Experiment. Math.}}
  \textbf{\bibinfo{volume}{16}}~(\bibinfo{number}{2}),
  \bibinfo{pages}{167--179} (\bibinfo{year}{2007}),
  \eprint{{arXiv:math.GT/0506295}}.

\bibitem[{Zhirov(1995)}]{Zhirov1995}
\bibinfo{author}{A.~Y. Zhirov}, \enquote{\bibinfo{title}{On the minimum
  dilation of pseudo-{A}nosov diffeomorphisms of a double torus},}
  \emph{\bibinfo{journal}{Russ. Math. Surv.}}
  \textbf{\bibinfo{volume}{50}}~(\bibinfo{number}{1}),
  \bibinfo{pages}{223--224} (\bibinfo{year}{1995}).

\bibitem[{Leininger(2003)}]{Leininger2004}
\bibinfo{author}{C.~J. Leininger}, \enquote{\bibinfo{title}{On groups generated
  by two positive multi-twists: {T}eichm\"{u}ller curves and {L}ehmer's
  number},} \emph{\bibinfo{journal}{Geom. Topol.}}
  \textbf{\bibinfo{volume}{8}}, \bibinfo{pages}{1301--1359}
  (\bibinfo{year}{2003}).

\bibitem[{Hironaka(2009)}]{Hironaka_WhatIs}
\bibinfo{author}{E.~Hironaka}, \enquote{\bibinfo{title}{What is\dots {L}ehmer's
  number?}} \emph{\bibinfo{journal}{NAMS}}
  \textbf{\bibinfo{volume}{56}}~(\bibinfo{number}{3}),
  \bibinfo{pages}{374--375} (\bibinfo{year}{2009}).

\bibitem[{Boyd(1980)}]{Boyd1980}
\bibinfo{author}{D.~W.~Boyd}, \enquote{\bibinfo{title}{Polynomials
    having small measure},} \emph{\bibinfo{journal}{Math. Comput.}}
  \textbf{\bibinfo{volume}{35}}~(\bibinfo{number}{152}),
  \bibinfo{pages}{1361--1377} (\bibinfo{year}{1980}).

\bibitem[{Penner and Harer(1991)}]{PennerHarer}
\bibinfo{author}{R.~C. Penner} and \bibinfo{author}{J.~L. Harer},
  \emph{\bibinfo{title}{Combinatorics of Train Tracks}}, number
  \bibinfo{number}{125} in \bibinfo{series}{Annals of Mathematics Studies}
  (\bibinfo{publisher}{Princeton University Press},
  \bibinfo{address}{Princeton, NJ}, \bibinfo{year}{1991}).

\end{thebibliography}

\end{document}